\documentclass[12pt]{amsart}
\begin{document}

\newcommand{\ppsi}{{\tilde{\psi}}}
\newcommand{\C}{{\mathbb C}}
\newcommand{\Q}{{\mathbb Q}}
\newcommand{\R}{{\mathbb R}}
\newcommand{\Z}{{\mathbb Z}}
\newcommand{\arctanh}{{\rm arctanh}}
\newcommand{\halfi}{{\textstyle{\frac{1}{2i}}}}
\newcommand{\half}{{\textstyle{\frac{1}{2}}}}
\newcommand{\PGl}{{\rm PGl}}

\title{On the visual appearance of relativistic objects} 
\author{Jack Morava}
\address{Department of Mathematics, Johns Hopkins University,
Baltimore, Maryland 21218}
\email{jack@math.jhu.edu}
\thanks{The author was supported in part by the NSF}
\date {8 May 2008}
\begin{abstract}{It has been almost half a century since the realization [1,3] that an object moving at
relativistic speeds (and observed by light reflected from some point source) is seen not as
squashed (as a naive interpretation of the Lorentz-Fitzgerald contraction would suggest) but
rather as {\bf rotated}, through an angle dependent on its velocity and direction of motion. I will
try to show here that this subject continues to be worth exploring.\medskip

\noindent
The author asserts his moral right to remuneration for any applications of these ideas to computer games or
big-budget sci-fi movies.} \end{abstract}

\maketitle

\noindent
{\bf 1} There is a detailed account of this subject in Taylor's textbook [2], which I will take as point 
of departure; but it will simplify some things to use notation slightly different from his, as follows: 
We observe an object moving across our field of view at constant velocity $v$, illuminated by some fixed point
light source. Choose coordinates in which we observers are situated at the origin, and such that
the path of the observed object is a line in the upper half of the $(x,y)$-plane, parallel to the 
$x$-axis (as in Taylor's Fig. X-5 p. 352). Let $\ppsi$ denote the angle between our line of sight to the
object, and its path, normalized so that $\ppsi = 0$ when the object crosses the $y$-axis; thus 
\[
\ppsi = \psi - \pi/2 \;,
\]
where $\psi$ is the angle of observation used in Taylor's calculations. \bigskip

\noindent
A relativistic ray-tracing argument now shows that the observed object will appear to us as 
rotated counterclockwise through an angle $\phi$, satisfying the equation [Taylor X-12 p. 356]
\[
\cos(\phi + \psi) = \frac{\cos \psi - v}{1 - v \cos \psi}
\]
(with the velocity $v$ measured in units such that the speed of light $c = 1$). \bigskip

\noindent
{\bf 2} Let 
\[
\lambda : S^1 = \R/2\pi \Z \rightarrow [-\infty,+\infty] 
\]
denote the function $x \mapsto \arctanh \sin x$: this is the inverse to the function studied by C.
Gudermann (1798 - 1852). It has many other representations, eg
\[
\lambda(x) = \log|\tan x + \sec x| = \frac{1}{2} \log \Big| \frac{1 + \sin x}{1 - \sin x}\Big| \;,
\]
and is antiperiodic: $\lambda(x + \pi) = - \lambda(x)$. It will often be convenient here to regard
it as defined on the interval $[-\pi,+\pi]$. Its derivative, as industrious students of 
elementary calculus learn, is $\sec x$, from which the power series representation
\[
\lambda(x) = \sum_{n \geq 0} (-1)^n E_n \; \frac{x^{2n+1}}{(2n+1)!}
\]
is easily deduced. [The Euler numbers $E_n$ are entry {\tt A000364} in Sloane's online
database of integer sequences.] \bigskip

\noindent
{\bf Definition} The two-variable formal power series
\[
X+_M Y := \lambda^{-1}(\lambda(X) + \lambda(Y)) = X + Y + \cdots \in \Q[[X,Y]]
\]
is a one-dimensional formal group law, with $\lambda$ as its logarithm.\bigskip

\noindent
This group law is closely related to Mercator's projection, discussed below. If we write 
\[
\nu = \arcsin v \;,
\]
then we can state the \bigskip

\noindent
{\bf Proposition:} The observed rotation $\phi$ satisfies the equation
\[
\phi + \ppsi = \nu +_M \ppsi \;.
\] 
\bigskip

\noindent
{\sc Proof:} We can restate Taylor's formula as
\[
\cos (\phi + \ppsi + \pi/2) = \frac{\cos(\ppsi + \pi/2) - \sin \nu}{1 - \sin \nu \cos 
(\ppsi + \pi/2)} \;,
\]
ie as 
\[
\sin (\phi + \ppsi) = \frac{\sin \ppsi + \sin \nu}{1 + \sin \nu \sin \ppsi} \;.
\]
\medskip

\noindent
On the other hand, the proposition asserts that 
\[
\lambda(\phi + \ppsi) = \arctanh \sin(\phi + \ppsi)
\]
equals 
\[
\lambda(\phi) + \lambda(\nu) = \arctanh \sin \ppsi + \arctanh \sin \nu \;.
\]
\medskip

\noindent
Taking hyperbolic tangent of both sides, and using the addition formula for that function, gives
\[
\sin(\phi + \ppsi) = \tanh \Bigl[ \arctanh \sin \ppsi + \arctanh \; v \Bigr] = 
\frac{\sin \ppsi + v }{1 + v \sin \ppsi}
\]
as claimed. \bigskip

\noindent
{\bf 3} Mercator's projection sends a point on the sphere with latitude $\phi$ to a point in the
plane with $y$-coordinate $\lambda(\phi)$; this is essentially just the logarithm of stereographic
projection. The formal group law defined above thus combines the line-of-sight angle $\ppsi$ and 
the relativistic velocity angle $\nu$ by sending them separately to their Mercator projections 
to the line, adds them as real numbers, and converts their sum back to an angle. In particular, 
when $\ppsi = 0$ (the moving object is at its closest approach) the object is seen as rotated 
through the angle $\arcsin v$. That relativistic geometry is somehow a complex (Wick) rotation 
of Euclidean geometry is well-known, but this seems to be a very explicit form of that 
correspondence. \bigskip

\noindent
It is easy to check that the derivative of 
\[
\lambda^{-1}(x) = \arcsin \tanh x
\]
is the hyperbolic secant. This suggests that, remarkably enough,
\[
\lambda^{-1}(x) = -i \lambda (ix) \;,
\]
which is also easily verified. In other words, if we write $\Lambda(x) = \lambda(ix)$, then 
\[
(\Lambda \circ \Lambda)(X) = - \; X 
\]
as formal power series.\bigskip

\noindent
Now $\sin x = j(\exp (ix))$ with $j(z) = \halfi (z - z^{-1})$ and $\tanh x =
k(\exp(2x))$ with 
\[
k(z) = \frac{z - 1}{z + 1} = 
\left[\begin{array}{cc}
        1 & -1 \\
        1 & +1 \end{array}\right] (z) \;,
\]
so $\lambda(x) = \half \log (k^{-1} \circ j)(\exp(ix)) = \log ([M](\exp(ix)))$, where $[M] 
\in \PGl_2(\C)$ is the element of order four represented by 

\newpage

\[
\left[\begin{array}{cc}
         1 & i \\
         i & 1  \end{array}\right] = 
\left[\begin{array}{cc}
         0 & 1 \\
         i & 0   \end{array}\right] \cdot \left[\begin{array}{cc}
                                                  1 & - i \\
                                                  1 & + i \end{array}\right] 
\;.
\] \bigskip

\noindent
The matrix on the right represents the Cayley transform
\[
z \mapsto C(z) =  \frac{z-i}{z+i}
\]
which, on the unit circle, is essentially just stereographic projection onto
the imaginary axis; thus 
\[
\lambda(x) = - \log iC(e^{ix}) \;.
\]
It follows that
\[
X +_M Y \equiv i \log \frac{\cos \half(X-Y) - i \sin \half(X+Y)}{\cos \half(X-Y) + i \sin \half(X+Y)} \; \; 
({\rm modulo} \; 2\pi)
\]
extends to a continuous map
\[
S^1 \times S^1 - \{(\pm \half \pi, \mp \half \pi) \} \ni (X,Y) \mapsto X +_M Y \in S^1 
\]
from the twice-punctured torus to the circle. \bigskip

\noindent
I'd like to thank the students in the 2004 undergraduate relativity course at JHU for their interest
and encouragement, and Claude LeBrun for helpful correspondence about the history of this question. \bigskip

\bibliographystyle{amsplain}

\end{document}